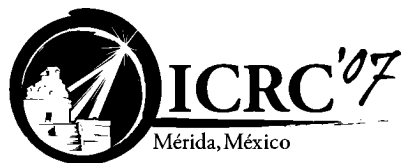

# The Galactic Gamma-Ray Club

ISABELLE GRENIER
*Laboratoire AIM, CEA/DSM-CNRS-Université Paris Diderot, Service d'Astrophysique, CEA Saclay, 91191 Gif/Yvette, France*
*isabelle.grenier@cea.fr*

**Abstract:** The exclusive Galactic $\gamma$-ray club has opened up to new members. Supernova remnants, pulsar wind nebulae, and massive binary systems hosting a compact object have recently joined the young pulsars as firmly established sources of $\gamma$ rays in the Milky Way. Massive young stellar clusters are on the waiting list to join the club. Only the fine imaging recently obtained at TeV energies could resolve specific sources. The samples are sparse, but raise exciting questions. The jet or pulsar-wind origin of the emission in binaries has been hotly debated, but it seems that both types of systems have been recently detected. The nature of the radiation in shock accelerators is still questioned: do nuclei contribute a lot, a little, or not to the $\gamma$ rays and what energy do they carry away from the shock budget? The acceleration process and the structural evolution of the pulsar winds are still uncertain. The magnetic field distribution in all these systems is a key, but poorly constrained, ingredient to model the multi-wavelength data, particle transport and electron ageing. It must, however, be determined in order to efficiently probe particle distributions and the acceleration mechanisms. The source samples soon to be expected from GLAST and the Cherenkov telescopes should bring new valuable test cases and they will, for the first time, shed statistical light on the collective behaviour of these different types of accelerators.

## GeV and TeV gamma-ray sources

Major advances in our understanding of the universe have often come from improving angular resolution at all wavelengths. The prowess of achieving several arc-minute resolution at TeV energies with the Cherenkov telescopes, and soon at GeV energies with GLAST, indeed opens a new era in $\gamma$-ray astronomy. The very secluded club of identified $\gamma$-ray sources, which has only accepted young pulsars and blazars for decades, has recently expanded to supernova remnants, pulsar wind nebulae, and $\gamma$-ray binaries. The high-energy facets of these objects of course raise new and exciting questions that I will try to briefly review. Yet, such a resolution does not compare with the imaging capabilities at lower wavelengths and deciding between true source identification and mere spatial coincidence in the crowded environments along the Milky Way will still be a key issue for years to come.

Source detection at GeV energies, unlike at TeV energies, has to fight against the intense and highly structured interstellar background that results from cosmic-ray interactions in the interstellar gas and soft radiation field. This background is of particular interest to the ICRC scientists because it probes the cosmic-ray density through the Galaxy, yet, with respect to point sources, its imperfect modelling induces systematic errors in source significance, flux, location, and spatial extension. Figure 1a illustrates the difference between the 3rd EGRET (3EG) catalogue [1] and the revised one presented at the conference [2]. The use of nine years of EGRET data instead of four at the time of the 3EG catalogue has allowed detection of 31 new sources, but the fact that as many as 107 former sources have not been confirmed in the new analysis is primarily due to the improved background that includes new CO clouds at medium latitudes, new HI maps corrected for stray radiation, and new maps for the dark gas phase in the nearby clouds. The latter is dark in the sense that it is not properly traced by HI and CO emission. This gas phase has been found in the nearby Gould Belt clouds [3]. It appears as an excess of both dust emission and diffuse $\gamma$ radiation over what the HI and CO data should yield. Both excesses are strongly correlated and form spatially coherent structures. So, both dust and cosmic rays trace some 'dark' gas that forms an extended layer at the transition between the dense CO cores and the outer HI envelope of the clouds. The dark gas column-densities compare with that measured in HI and CO, so it provides both $\gamma$-ray intensity and structure that were not accounted for in the earlier background models. Because source detection methods in $\gamma$ rays search for significant and point-like photon excesses above the *predicted* background, an ensemble of point sources with the wide EGRET PSF would compensate for the missing cloud structures and yield an excellent fit to the data. These 3EG sources have not been confirmed in the new analysis; they consist of only six sources with a likely AGN counterpart and 101 unidentified sources, in particular the faint and persistent ones that have long been spatially associated with the Gould Belt [4, 5, 6]. The improved GLAST resolution [7] will soon provide much more accurate maps of the diffuse interstellar emission, but the same uncertainties in gas tracers and in the cosmic-ray flux pervading the distant clouds will still impact source detection and localization near the detection threshold. At TeV energies, the interstellar intensity falls off more rapidly than most source spectra.

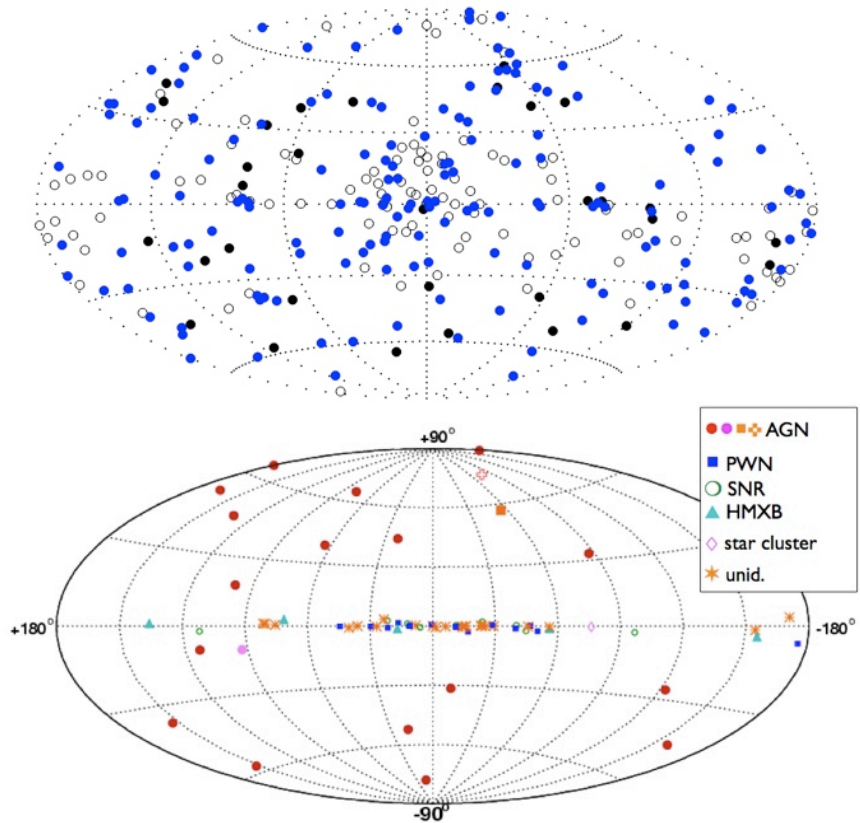

Figure 1: Distribution in Galactic coordinates of (a) the revised EGRET point sources detected at energies above 100 MeV on top of an improved interstellar emission model and using 9 years of EGRET data. Blue dots, open circles, and black dots respectively mark the confirmed 3EG sources, the unconfirmed ones and new sources; (b) the sources currently detected at TeV energies by the Cherenkov telescopes (courtesy of Robert Wagner).

This is why we presently detect more TeV than GeV sources at low latitude, as illustrated in Figure 1b. The concentration of sources at |l| < 60° is due to the coverage of the HESS survey.

## Irradiated clouds

It is, however, striking that the vast majority of TeV sources have radial extensions of several arc minutes that correspond to diameters of 20 or 30 parsecs in the inner spiral arms and molecular ring of the Milky Way. Pulsar wind nebulae have been seen as extended sources of TeV radiation and will be reviewed below, yet an exciting alternative is that some of these sources trace where the cosmic-ray sources dwell, in other words that they are due to pockets of enhanced cosmic-ray flux diffusing away from their source(s) and irradiating the gas on their way. The HESS J1800-240 A and B sources may serve as useful examples [8]. They correlate well with the integrated CO intensity map of massive $(0.5-1.5 \ 10^5 \ M_\odot)$ clouds at a distance of 2 and 4 kpc in the Sagitttarius and Scutum-Crux arms. The TeV flux requires 10 to 30 times the local cosmic-ray density if it is due to neutral pion decay. This is much more than the modest increase in cosmic-ray density expected in the inner Galaxy from large-scale studies of the interstellar GeV emission [9]. The increase with respect to the flux at the solar circle does not exceed a factor of 1.5 or 2. Another source in this direction, HESS J1801-233, correlates with CO intensity. It coincides with the rim of the W 28 supernova remnant, but may not be related to it. The shock wave from W 28 is known to run into 2000 solar masses of molecular gas which could be heavily irradiated if acceleration at the shock manages to remain active in the dense medium or if particles accelerated earlier and trapped downstream still leak out of the remnant. Yet, the HESS source spatially overlaps with only half of the shocked gas. The other half exhibits almost as much mass, but is not detected. It is not obvious that the small mass deficit compared to the TeV-bright side can explain the lack of emission. The TeV source correlates with a foreground or background CO complex of about $2 \ 10^4 \ M_\odot$ which may or may not be spatially related to the shocked clouds. It is difficult to tell in this crowded direction. Again, 10 to 30 times the local cosmic-ray density is needed to explain the flux in terms of pion decay in this cloud. It could be provided by W 28 or by other sources embedded in the cloud.

The three HESS sources associated with CO exhibit equivalent spectra, with soft photon indices of 2.5-2.7 typical of interstellar emission. When protons of energy $E_p$ diffuse for a time $\tau$ away from their source, with an interstellar diffusion coefficient $D \approx 10^{24} \ (E_p/10 \ GeV)^{0.6} \ m^2 \ s^{-1}$, the pion-decay spectrum breaks at an energy $E_\gamma \approx 0.17 \ E_p$ that shifts with distance L from the source as $L^2 \approx 6 \ D(E_p).\tau$ [10]. Lower energy particles cannot reach that far in the same amount of time, so the spectrum drops at lower energies. The spectrum above the broad peak falls as $E^{-2.6}$ for an $E^{-2}$ source injection spectrum. The flux also peaks around this characteristic timescale $\tau(L, E)$ since spherical expansion dilutes the flux on longer time scales (in the same energy band). The clouds of $10^4$ to $10^5$ solar masses that we are discussing have typical sizes of 20 to 60 pc, respectively, near virial equilibrium. The soft $E^{-2.7-2.5}$ spectra seen by HESS require a peak proton energy of 600 GeV or less, that can be seen out to 50 pc after a thousand years of diffusion. The flux estimates derived by Aharonian & Atoyan [10] agree reasonably well with the observed ones. More distant clouds would exhibit too hard spectra and undetectable fluxes at TeV energies. The same 50 pc cloud complex would yield a flux detectable by GLAST after 50 kyr, with a flat energy spectrum near the peak that would allow separation from the bulk of the diffuse interstellar emission. It should be noted, however, that the diffusion properties in the interstellar medium are not well characterized, especially in dense environments where lower diffusion coefficients may prevail. For a 10 times lower coefficient, one would have to wait for 10 kyr for the higher energy particles to diffuse out to get a soft enough TeV spectrum in the 50 pc large cloud, but not too long because the flux drops by an order of magnitude after 50 kyr [11]. Establishing several examples of 'over-irradiated' clouds at GeV and TeV energies would therefore prove very useful to explore the diffusion properties as well as to locate where the cosmic-ray sources live.

Another example of cloud irradiation has been proposed toward the Galactic centre [12] where the correlation between the molecular gas distribution (as traced by CS) and the TeV intensity profile may suggest an overabundance of freshly accelerated cosmic rays. Given the turmoil in the inner 150 pc of the Milky Way, it is unfortunately one of the most difficult places to interpret spatial distributions. Given the tortured structure of the magnetic field and its possible strength of 100 nT to account for the bright radio arcs, one cannot use the local value of the diffusion coefficient. The latter should also largely differ in and out of the plane. One may ask how the particles can efficiently cross the radio arc region and numerous other magnetic threads that stretch perpendicular to the plane between the central TeV source assumed by the authors and the peaks detected at l = ± 0.6°. An estimate of the fraction of particles that escape out vertically would be useful.

A second problem is that the secondary electron-positron flux produced in the hadronic interactions fails by nearly a factor 100 to mach the radio emission found in Sgr B2. It does not explain either the unusually high ionization level measured in the cloud (~ 4 $10^{-16}$ s$^{-1}$) [13]. A pure inverse Compton origin of the TeV data is difficult to reconcile with the X-ray and radio fluxes in the cloud and the required magnetic field is too low [13]. On the other hand, the distribution of the TeV emission in the whole region follows that of the fluorescent 6.4 keV line from neutral iron which is probably excited in the clouds by very low energy cosmic rays [14]. So these clouds, unlike others at |ℓ| > 1°, appear to harbour an excess of long-lived, low-energy cosmic rays as well as fresh high-energy ones. A passing wave of particles produced by Sgr A East or another central supernova remnant of the past millennia is not the end of the story.

## Gamma-ray champagne bubbles?

Alternatively, the TeV intensity around Sgr A* correlates with the position of the very young and incredibly active star-forming regions of the Arches, Quintuplet, and Sgr B2 clusters, as well as the less prominent, but still very active, Sgr B1 and C regions. TeV radiation has also been detected toward two other young and massive stellar clusters: by HESS toward Westerlund 2 in the giant RCW49 HII region [15], and by Milagro and MAGIC on the edge of Cygnus OB2 [16, 17]. These sources, HESS J1023-575 and TeV J2032+4130, have not been identified with other likely γ-ray emitters such as supernova remnants, pulsars or their wind nebulae. They do not coincide with EGRET sources. These spatial coincidences raise the interesting possibility of efficient particle acceleration in these young environments.

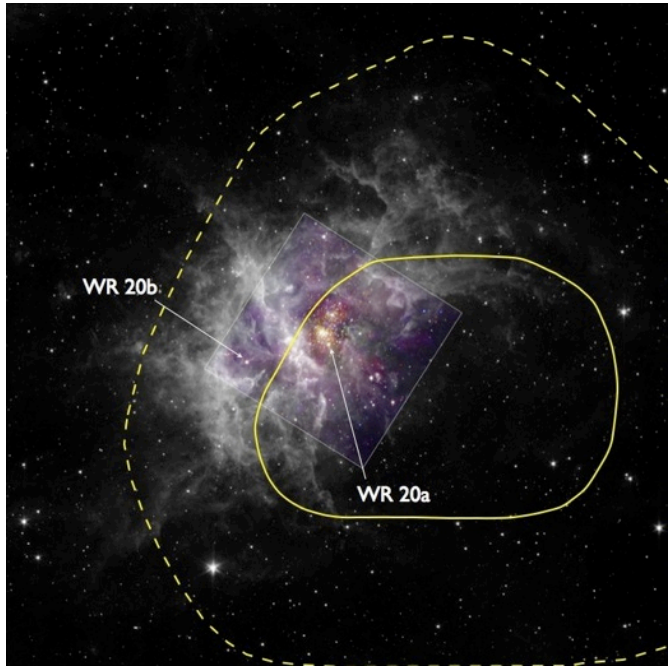

Coincidences between COS-B sub-GeV sources and OB associations, together with anomalies in the cosmic-ray composition, had prompted the idea of SNOBs [18] in which ions would be first injected and accelerated by the supersonic winds of massive stars and then by the shock wave of a nearby supernova remnant. The recent measurements by ACE of isotopic ratios in the local cosmic rays point to an acceleration site inside OB associations to explain the presence of about 20 % Wolf-Rayet material in the composition [19]. The lack of $^{59}$Ni also suggests a period of at least $10^5$ years between nucleosynthesis and acceleration [20], supporting the two-step scenario of SNOBs. The cloud complex that has given birth to the cluster provides targets for hadronic interactions to shine in γ rays. The power released by OB associations in the combined form of stellar winds and supernova shocks is large enough to sustain the required γ-ray luminosities for standard mechanical-to-cosmic-ray energy conversion efficiencies of a few percent. EGRET, however, has seen only few sources toward OB associations (in the Cygnus region, Car OB 1b and 2, Sco OB2d, and Gem OB1) and they may only be chance coincidences in these crowded directions.

Potential associations with TeV sources appear to be more promising, but different because the related OB clusters are rather unique in stellar content and youth [21, 22, 23, 24].

Figure 2: significance contours of the HESS J1023-575 source (7 and 9 σ as dashed and solid curves, respectively) overlaid on the Spitzer and Chandra images of Westerlund 2 and its HII region RCW 49. The PAH infrared emission (in grey) outlines the cloud highly perturbed response to the ionization fronts and winds from the very massive stars. The X rays (in colour) reveal the compact and massive stellar cluster that powers the intense activity in this highly obscured region.

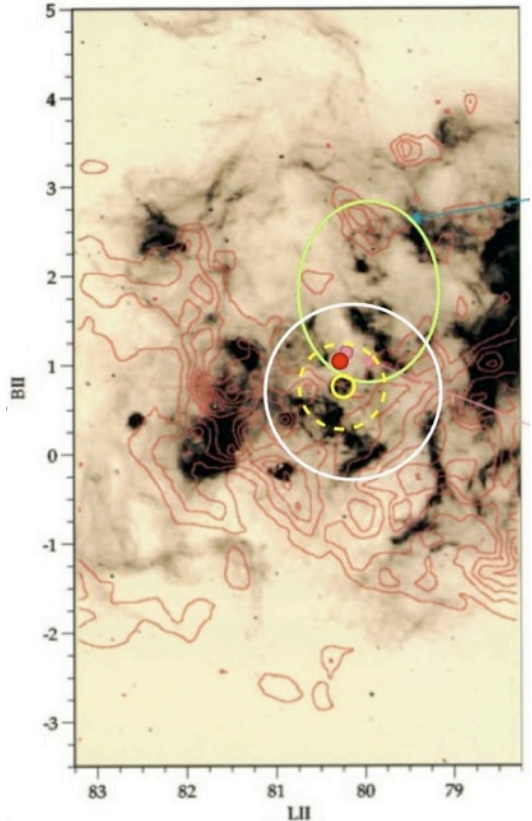

Figure 3: location and extent of the TeV J2032+4130 source (red dot, as seen by MAGIC [17]) with respect to the massive Cyg OB2 cluster (white circle, [23]) breaking out in a CO shell (green ellipse). The parent cloud and ionized gas are shown as CO contours (red) overlaid on the 1420 MHz intensity map [25]). The EGRET source EGR J2033+41 is centred within the yellow error circle, but any extended source within the dashed yellow circle would not be resolved by EGRET.

They are referred to as 'super star clusters'. They result from extraordinary bursts of star formation that have converted $10^{3.7}$ to $10^{4.6}$ solar masses of gas in stars heavier than the Sun. They are extremely rich in massive stars (e.g. 150 O stars in the Arches, $120 \pm 20$ O stars in Cyg OB2, down to > 12 O stars in the more modest Westerlund 2 cluster). The huge ionizing photon fluxes of $10^{50.8}$ to $10^{51.6}$ s$^{-1}$ have carved out wide HII regions around them. These photons can efficiently produce pairs on the TeV $\gamma$ rays, but one can easily check that, given the cluster sizes, the photon densities are not lethal to $\gamma$ rays except within light minutes of one of the massive members [26]. The inverse Compton scattering of all the starlight can double the amount of interstellar GeV $\gamma$ rays in the cluster vicinity. In the case of Cyg OB2, it would form a 1° wide source of $\gamma$ rays with a flux near the threshold of the GLAST telescope [27]. All these clusters have manufactured giant stars of 85 M$\odot$ in Westerlund 2 and over 100 M$\odot$ in the Arches and Quintuplet. The large number of LBV and WR members that are still alive attests of their youth (5% of all known Galactic WR in the Arches, > 17 in the Quintuplet, 3 in Cyg OB2, and 2 in Westerlund 2). Ages of $2.5 \pm 0.5$, 2, and 2-3 Myr have been derived for the Arches, Cyg OB2, and Westerlund 2, respectively [22, 28, 29]. Little is known from Sgr B2 since the stars still hide in their parent cloud, but the presence of 50 compact HII regions strongly suggests it is another Arches cluster in the making. The Quintuplet is slightly more evolved, with an age of $4 \pm 1$ Myr.

If these systems turn out to be responsible for the TeV emission, their extreme youth calls for another scenario than the more evolved case of SNOBs because of the lack of supernovae. Even if remnants could easily hide in the intense free-free radio emission, the clusters are too young to have produced several supernovae. Mechanically, these young systems are almost as powerful as SNOBs. Cesarsky & Montmerle [30] have estimated that stellar winds from WR and O stars dominate the cluster power during the first few million years, especially in the case of massive associations with stars heavier than 30 M$\odot$ as observed in the super-star clusters. The power exceeds $10^{31}$ W if the cluster hosts 40 stars with mass-losses > $10^{-5}$ M$\odot$ yr$^{-1}$ and terminal velocities of order 2500 km/s. Supernovae hardly take over the overall budget when the cluster is 5 or 6 million years old. Whereas Fermi acceleration by multiple random shocks from supernova remnants in more evolved OB associations has been studied to account for cosmic-ray energies and composition near or beyond the knee [31], the shock sizes, velocities, and separation in space and time do not apply to the denser, more turbulent case of the colliding winds from tens of massive stars in the compact configuration of the super clusters. Losses in the dense bubble shells also limit the acceleration. Figure 2 shows how intensely the stars have ionized and sculpted the gas around them in Westerlund 2. A ridge of radio emission follows the rim of the bubble centred on the main cluster and along its interface with the bubble blown by WR 20b, but it is dominated by free-free emission. Particle acceleration has been observed as synchrotron radio emission in the colliding winds of massive stars within binary systems. It should yield detectable fluxes of $\gamma$ rays, if not at TeV energies because of the heavy toll of pair absorption on the intense stellar radiation field, at least to tens of GeV energies [32]. However, the large spatial extent of HESS J1023-575 toward Westerlund 2 (25 pc in radius at 8 kpc) is not compatible with an origin of the emission in the colliding winds of the WR 20a binary. The whole Westerlund 2 cluster is also too compact to explain the large TeV source. A more diffuse cause, perhaps in the collective action of separate winds, is needed. This is also true for the more compact case of the TeV J2032+4130 source (4.4 pc in size at 1.5 kpc) which is seen toward a sub-group of 10 or more O stars of the Cyg OB2 cluster. The position of the EGRET EGR J2033+41 source (consistent with 3EG J2033+4118) is well centred on the dense core of the cluster and it could encompass emission from the entire system because of the wide EGRET point-spread-function. The massive eclipsing binary V729 Cyg in Cyg OB2 cannot account for the observed luminosity [33].

Interestingly, the γ-ray sources toward Westerlund 2 and Cyg OB2 are found where the hot bubble of the main HII region breaks free into the low density medium, blowing away the ambient gas in a champagne flow that sweeps the magnetic field lines away with it in a mushroom-like configuration. Adiabatic losses in the expanding flow must be severe, but particles can stream along the field lines and flow back to the edge of the cloud, or they can diffuse along the edge of the champagne flow to produce γ rays [34]. The problem is to accelerate cosmic rays from the bubbling inside the champagne bottle, within the HII bubble, to begin with. This γ-ray champagne bubble scenario, as well as the possible connection between young massive clusters and γ-ray sources has yet to be verified observationally in other examples, taking advantage of the upcoming surveys of the Galactic plane by GLAST and the Cherenkov telescopes and combining their data to constrain the high-energy spectrum. Having missed until recently the importance of Cyg OB2 at a mere distance of 1.5 kpc reminds us that super star clusters may have escaped radio searches in the Galaxy.

## Cosmic-ray acceleration in supernova remnants

Collisionless shocks in supernova remnants are thought to produce the bulk of the cosmic rays up to the knee. They are numerous and powerful enough to sustain the total cosmic-ray power in the Galaxy. Synchrotron X rays also indicate that the forward shock effectively accelerates electrons to tens of TeV (see below). Yet, the accelerated ions are very elusive and they have not been firmly observed so far. Strong, but indirect, evidence of their presence has come from several important clues in the X-ray images. The thinness of the syn-chrotron filaments observed behind the forward shock of young remnants (such as Cas A, Kepler, Tycho, and SN 1006) imply very severe losses from large magnetic field strengths, well in excess of the compressed interstellar field ([35] and references therein). The electrons rapidly cool down and their radiation shifts to UV energies, below the spectral window of the X-ray telescope. It has been proposed that the magnetic amplification is due to cosmic-ray streaming upstream [36, 37]. Another clue comes from the X-ray morphology of these remnants. All models of diffusive shock acceleration predict that ions receive far more energy than electrons. When this energy drain becomes significant, the gas hydrodynamics and shock profile are modified. The shocked gas becomes more compressible and it piles up in a much thinner shell outside the contact discontinuity. Instead of the classical compression ratio r = 4, average values of $6 \le r \le 8$ are predicted after several centuries of acceleration activity. The post-shock temperature thus drops by an order of magnitude or more compared to the classical case [38]. Several observations support this possibility. A modified shock can reconcile the unusually low electron temperature measured in the young 1E0102.2-7219 remnant with its large shock velocity [39]. Detailed spectro-imaging of Tycho and Kepler also shows that the forward shock is twice closer to the contact discontinuity than one would expect when the shock is not disturbed by accelerated ions [40]. These are strong indications of efficient acceleration. The case of Cas A is less conclusive. A wide shell is seen between the shock and the ejecta. It may be due to inefficient acceleration, despite the large magnetic amplification implied by thin X-ray filaments, or to the remnant expansion in the wind of its progenitor [40].

Three remnants, G347.3-0.5, Vela Jr, and RCW 86, have now been imaged at TeV energies [41, 42, 43]. G347.3-0.5 and RCW 86 may be the remnants of the historical supernovae SN 393 and SN 185. The three objects share many common traits. Their large size (11 pc at 1.3 kpc, 17 pc at 1 kpc, and 18 pc at 2.8 kpc, respectively), their large ratio of synchrotron to thermal X-ray flux, together with their weak radio emission suggest that they have expanded at high speed in a low-density medium, possibly a stellar wind bubble or an OB cavity. So, they would be near the end of the free expansion phase or in early Sedov stage. Part of their shells has reached a denser environment a few centuries ago (to the west and southwest of G347.3-0.5 [44] and southwest of RCW 86 [45]). Their non-thermal X-ray filaments are quite broader than those of the younger remnants quoted above, thus implying a lower degree of magnetic amplification. The TeV flux appears well correlated with the X-ray synchrotron one. Synchrotron losses dominate over inverse Compton ones in their spectral energy distributions (SEDs), even if the TeV emission is entirely electronic. It is my purpose here to grossly constrain the range of shock velocities ($v_{sh}$), post-shock magnetic fields ($B_d$), and diffusion coefficients at high energy that are consistent with the observed energies and flux ratios of the SED peaks, and with the observed width of the synchrotron and $\gamma$-ray profiles. It is also my purpose to convey that both possible origins for the $\gamma$ rays (pion-decay from in-situ ions or electrons up-scattering the cosmological microwave background (CMB) radiation and dust infrared emission) still face serious difficulties.

Following the formalism of Parizot et al. [35], I will assume that both the upstream and downstream diffusion coefficients D(E) scale as $k_0$ times the Bohm limit, so $D(E) = k_0 E/3eB$ for an electron or proton of energy E in a B field. The most efficient acceleration takes place when $k_0 = 1$. I also assume a mean compression ratio $r = 7$ and a ratio of downstream to upstream magnetic field strength $B_d/B_u = 0.83 \times r$ in the case of isotropic magnetic turbulence [46]. The synchrotron peak energy and filament width provides two constraints. Equating the 1st order Fermi acceleration timescale and the synchrotron loss timescale upstream and downstream yields an estimate of the maximum electron energy $E_{emax} \propto B_d^{-1/2} k_0^{-1/2} v_{sh}$. The synchrotron cut-off energy therefore scales as $h\nu_{syn\text{-}cut} \propto v_{sh}^2 k_0^{-1}$ and provides a first constraint between the shock velocity and $k_0$. It is displayed in the left plots of Figure 4 for the three remnants. Because of diffusion, advection, and synchrotron losses, the high-energy electron distribution falls off exponentially downstream and the synchrotron profile at a given X-ray frequency, $h\nu_X$, has a projected full width at half maximum $\Delta\theta_{syn}$ that follows equation 24 in [35]. The ratio $B_d^{3/2}/v_{sh}$ is thus constrained by the three observables $h\nu_{syn\text{-}cut}$, $h\nu_X$, and

$\Delta\theta_{syn}$. This relation is plotted as the blue line in the right plots of Figure 4. The cut-off energies being poorly constrained in the current SEDs, an uncertainty within a decade is quite plausible. It implies an equivalent uncertainty in $k_0$. Its impact is shown in Figure 4 as the blue shaded area. For the input parameters of the three remnants, one finds that the downstream field roughly scales with $h\nu_{syn\text{-}cut}^{-1/3}\Delta\theta_{syn}^{-2/3}$. This is why the constraint on $B_d$, displayed again as the blue shaded area, is more robust.

The $\gamma$ rays provide other constraints. The peak inverse Compton (IC) flux cannot exceed the peak $\gamma$-ray flux observed in the SED, so the magnetic energy density has a lower limit set by this flux ratio and the energy density of the ambient soft radiation field. This lower limit is met if one assumes the TeV data to arise from pure IC emission. The lower limit appears as the lower horizontal red line in Figure 4. I have used 0.26 MeV m$^{-3}$ for the cosmological background and 0.27 MeV m$^{-3}$ for the local density of of thermal emission from cold (24 K) dust for Vela Jr and G347.3-0.5 [47]. RCW 86 resides in the much brighter inner Galaxy, but in the quieter interarm region between the Carina and Scutum-Crux arms, so the ambient inrared density may be twice larger than the local value. If IC emission dominates over pion decay, the peak $\gamma$-ray energy, $h\nu_{\gamma cut}$, provides another relation between the downstream field, $h\nu_{syn\text{-}cut}$, $h\nu_{\gamma cut}$, and the soft photon energies ($E_{soft} = 0.66$ and 5.8 meV for the CMB and cold dust photons, respectively). It appears as the grey shaded area in Figure 4. In the same IC framework, the width of the TeV profile should not exceed the observed one. This provides another lower limit in the ($B_d$, $v_{sh}$) phase space that depends on $E_{soft}$, $h\nu_{syn\text{-}cut}$ and the intrinsic width $\Delta R_\gamma$ of the $\gamma$-ray shell as inferred from the observations at the energy $h\nu_\gamma$. This lower limit appears as the two broken red lines in Figure 4 (solid and dotted for the CMB and IR target photons, respectively). The blue lines and broken red lines change by only 25% when using the classical r = 4 compression ratio instead of r = 7. The observational constraints summarized in Table 1 were taken from the data in [35, 41, 42, 43, 45, and references therein]. In the absence of velocity measurements, one expects the shock speed to vary with the remnant radius R and age $\tau_{age}$ as $v_{sh} = 2R/3\tau_{age}$ during the free expansion phase. The velocity can be derived for G347.3-0.5 and RCW 86 if they are related to the historical supernovae. It is plotted as the vertical dashed line in Figure 4. On the other hand, we can set a maximum velocity by reasonably requiring the age of the three objects to be older than 1000 years. The upper limit is plotted as the solid vertical line in Figure 4.

| | G 347.3-0.5 | Vela Jr | RCW 86 |
|---|---|---|---|
| distance (kpc) | 1.3 | 0.2 or 1.0 | 2.8 |
| angular radius | 30' | 60' | 22' |
| projected FWHM of the X-ray filaments | 40" at 2 keV | 50" at 5 keV | 100" at 2 keV |
| intrinsic fractional width dR/R of the shell at 1 TeV | 45 % | 22.5 % | 50 % |
| synchrotron peak energy (keV) | ~ 1 | ~ 0.9 | ~ 0.15 |
| synchrotron peak energy flux (eV cm$^{-2}$ s$^{-1}$) | ~ 190 | ~ 100 | ~ 8 |
| inverse Compton peak energy (TeV) | ~ 5 | ~ 4 | ~ 1 |
| inverse Compton peak energy flux (eV cm$^{-2}$ s$^{-1}$) | ~ 19 | ~ 20 | ~ 4 |

Table 2: supernova remnant characteristics used to constrain the downstream magnetic field and shock velocity.

The results for G347.3-0.5, Vela Jr at 1 kpc, and RCW 86 show that their current shock speed should exceed 3800, 3600, and 1500 km/s, respectively, to accelerate electrons to the observed cut-off energies in X rays with the maximum efficiency ($k_0 = 1$, Bohm limit). Large field strengths, of order 10 nT, are also required to explain the thinness of the X-ray synchrotron filaments. These values are less subject to the uncertainty in the synchrotron cut-off energy. As expected, they are slightly lower than the fields strengths of 20 to 30 nT derived for the younger historical remnants [35].

Let us first examine the case of dominant IC emission at high energy. The large magnetic fields required by the X-ray filaments are consistent with the peak energy observed by HESS. The energetic electrons preferably up-scatter soft, CMB-like, photons which correspond to the lower end of the grey shaded region. The large magnetic fields are also consistent with the thickness of the TeV shell. IC emission also provides a natural explanation for the tight spatial correlation between the X-ray and TeV fluxes. Yet, the necessary large magnetic fields are not consistent with a dominant IC origin of the γ-ray flux. A 'pure' electronic origin requires the low fields of 1.5, 1.0, and 0.8 nT inferred from the peak-flux ratio between the X-ray and γ-ray components for the 3 remnants. Similar values have been found for G347.3-0.5 and Vela Jr by modelling the observed SEDs with a power-law distribution of electrons [41, 42]. These fields are, however, too low to explain the sharp synchrotron filaments and are only marginally consistent with the TeV shell thickness. One would need to increase the sub-mm part of the dust emission by two orders of magnitude to reconcile the dominant IC scenario and high magnetic fields. This option is unlikely. Another possibility is to suppose that the large magnetic fields are confined to sharp filaments at the shock and that they rapidly drop inward [48]. If so, the filaments would trace the magnetic radial profile rather than fast synchrotron cooling and the field strength $B_d$ would not be constrained by the blue areas in Figure 4. This option requires a careful modelling of the diffusion, advection, and radiation cooling to be tested against the observations because of the importance of the projection effects. Electron spectral hardening with energy, as expected from modified shocks, should also be taken into account. A convincing IC scenario would also need to explain why the synchrotron spectral index varies from 1.8 to 2.6 along the shell of G347.3-0.5, whereas the $\gamma$-ray index remains uniformly close to 2.0 [41, 44]. The large magnetic field helps in this matter since it implies X-ray electrons that have a few times more energy than the $\gamma$-ray emitting ones. So, the latter are closer to the cut-off energy (thus harder) than the X-ray ones that belong to the exponential cut-off part of the spectrum.

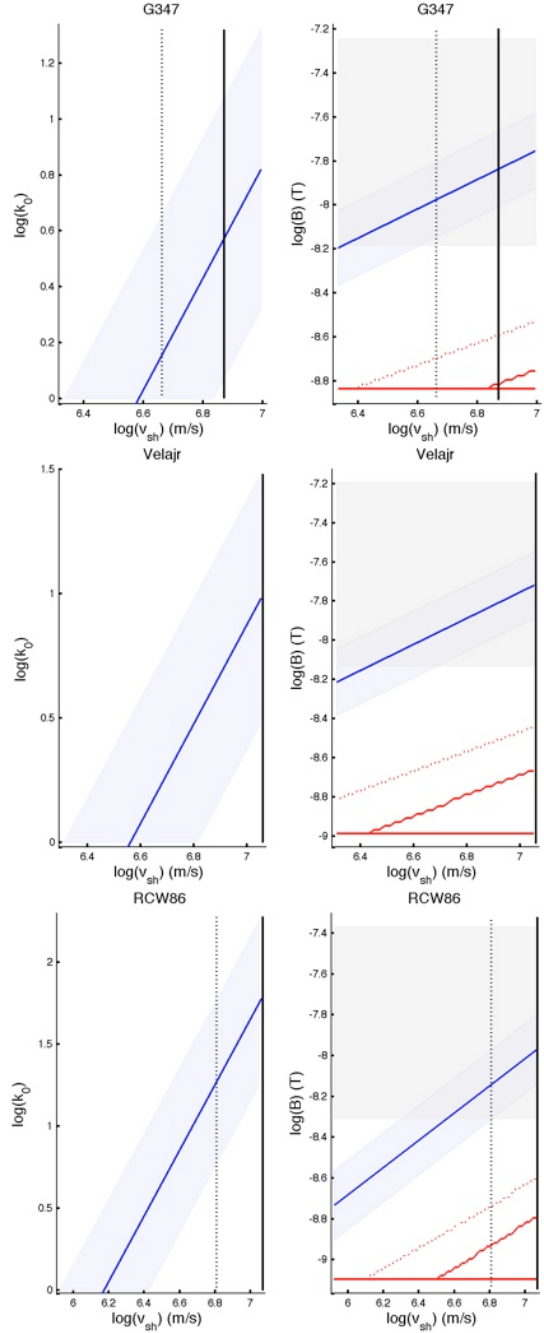

Figure 4: constraints on the diffusion coefficient ($k_0$ times the Bohm limit) and on the downstream magnetic field as a function of shock velocity for the three supernova remnants, G347.3-0.5, Vela Jr at 1 kpc, and RCW 86, resolved at TeV energies (see text for the explanation of the different lines and regions).

The nearby distance of 200 pc for Vela Jr does not yield a convincing case. The minimum shock velocity necessary to sustain the X-ray synchrotron properties is 1.6 times larger than the maximum value set by an age older than 1 kyr and the required field is very large ($B_d$ > 25 nT). It would completely rule out an IC scenario, as noted by the HESS group [42].

Following the constraints in Figure 4 and adopting downstream field strengths of 8, 8, and 6 nT as possible values for G347.3-0.5, Vela Jr (1 kpc), and RCW 86, respectively, we can infer maximum electron energies of 26, 24, and 11 TeV and synchrotron lifetimes of 100, 110, and 380 yr. The maximum proton energy is derived by equating the acceleration timescale and the remnant age. Choosing ages of 1620 and 1820 years from the historical supernovae and a comparable age of 2300 years for Vela Jr at 1 kpc that corresponds to a current speed of 5000 km/s in Figure 4, we find that protons can be accelerated to 400 TeV in G347.3-0.5, 500 TeV in Vela Jr, and 50 TeV in RCW 86. These values fall well below the knee energy despite the large magnetic amplification. The region explored by these high-energy protons in their random walk is found to be smaller than the observed TeV shell thickness.

The large magnetic amplification favours the interpretation of the $\gamma$-ray flux in terms of pion decay, but this scenario also faces important difficulties. Large gas densities between 1 and 2 cm$^{-3}$ are required to match the TeV flux. Yet, the absence of thermal emission in G347.3-0.5 sets an upper limit of 0.02 cm$^{-3}$ at 1 kpc [44]. This limit can be further lowered for an increased gas compressibility in a modified shock. A low density of 0.03 ff$^{-1/2}$ cm$^{-3}$ (for a filling factor ff) has also been found for Vela Jr at 1 kpc [49]. These values cannot be increased to meet the $\gamma$-ray requirement. Large densities (300 cm$^{-3}$) exist on the southwestern side of G347.3-0.5 where the clear correlation between the synchrotron flux and absorbing column-density indicates that the shock runs into dense clouds seen in CO. If protons are accelerated as well as electrons, having more high-energy protons and more targets to produce pions, one would expect the pion-decay flux to scale more or less quadratically with the ambient density. Yet, the $\gamma$-ray profile linearly follows the X-ray one, both radially and azimuthally. It does not exhibit an enhanced $\gamma$-ray to X-ray flux ratio toward the CO clouds [41]. The same difficulty is seen in RCW 86. The TeV map presented at the conference carefully avoids the southwestern region where the shock is known to have slowed down to 800 km/s inside a dense cloud and where ample thermal emission is detected [45]. In addition, the pion-decay modelling of the TeV emission in Vela Jr requires an unsually low electron to proton ratio (< 10$^{-4}$). Furthermore, the maximum proton energies derived from the amplified magnetic fields chosen above correspond to cut-off energies in the $\gamma$-ray spectrum of 70 TeV for G347.3-0.5 and 90 TeV for Vela Jr, at odds with the spectral break detected between 3.7 ± 1.0 TeV and 17.9 ± 3.3 TeV in G347.3-0.5 [41] and possibly several TeV for Vela Jr [42]. One would need to decrease the downstream field by an order of magnitude to explain these breaks from proton-proton collisions. It would then conflict with the observed synchrotron properties and IC emission would become dominant. The spectral break near 9 TeV inferred for RCW 86 would be consistent with the possible 5 TeV break reported at the conference. Finally, it has often been said that a flat spectrum below 1 TeV would sign pion-decay against IC emission, but the reader should be warned that the flattening at high energy of the electron spectra emerging from modified shocks, as well as various dosages of infrared photon targets can easily broaden the IC spectra and increase the overall IC flux for a given magnetic amplitude. Flattened spectra will certainly help to better fit the radio and X-ray data simultaneously, but they will provide less room for pion-decay emission.

In conclusion, whereas indirect evidence of ion acceleration exists in X rays from the modified shock thermodynamics, no clear picture has emerged yet from the beautiful $\gamma$-ray images of shell supernova remnants.

## Pulsar wind nebulae

The case of the HESS J1813-178 source brings forth the problem of separating emission from the supernova remnant and pulsar activities when the telescope resolution is limited. The source coincides with the radio shell of the G12.82-0.02 remnant and with a non-thermal X-ray nebula that probably traces the wind of a young and energetic pulsar inside the remnant, although searches for radio pulsations have failed so far [50]. Since the 1990s, when the GeV to TeV emission from the Crab nebula was successfully interpreted as synchro-self-Compton emission in the pulsar wind downstream of its terminal shock inside the remnant [51], and when several of us drew attention to the fact that a number of hard X-ray nebulae were seen inside the error boxes of unidentified, possibly variable, in EGRET sources (e. g. [52, 53]), pulsar wind nebulae have been firmly established as sources of TeV emission. The identification is based on the correlation between the X-ray and γ-ray images of the diffuse emission. In several cases, the TeV image extends further away from the pulsar, as around PSR B1823-13 (HESS J1825-137, [54]), in the long Vela X tail [55], or along the jet of PSR B1509-58 [56]. For more compact sources, unresolved by HESS, such as in G21.5-0.9, G0.9+0.1, and Kes 75, the proposed identification is based on the lack of non-thermal X-ray emission from the shell or on the weakness of the magnetic field required to explain the TeV flux from electrons accelerated at the shell [57, 58].

The number of pulsar wind nebulae emitting synchrotron X rays has grown rapidly in the past few years. Thanks to the high-resolution images provided by XMM-Newton and Chandra, many have been resolved into a large variety of shapes. The innermost regions of the wind are dominated by polar jets and a toroidal wind that results from the winding of the neutron star magnetic field. A rather complex standing shock forms as the wind decelerates to match the boundary condition imposed by the external medium (the pressure of the supernova ejecta or the ram pressure of the ambient medium if the pulsar motion has become supersonic inside or outside the remnant). The downstream equatorial flow expands laterally and turns back against the external medium to fill intermediate latitudes and form an elongated bubble, stretched along the pulsar spin axis and jets [59]. The synchrotron emitting part of the wind is confined between the termination shock and the bubble outer boundary since it is assumed that the upstream charges flow along with the frozen-in field and do not radiate. Doppler boosting of the synchrotron intensity is important for the innermost part of the wind, as in the Crab wisps, but not for the rest of the nebula [59]. The wind electrons can efficiently upscatter soft photons to produce TeV γ rays. In the Crab nebula, the average synchrotron infrared and optical energy density in the bubble exceeds the other ambient radiation fields and the modelling of the multi-wavelength spectrum indeed finds that synchro-self-Compton emission dominates in γ rays [51, 60]. The situation appears to be different for the other nebulae seen in γ rays. The second most luminous one after the Crab is in G21.5-0.9. Assuming that the sub-eV synchrotron energy flux is comparable to the observed X-ray one (for a flat SED), and considering the size of the synchrotron bubble (0.65', [61]), one finds an average energy density 7 times lower than that of the cosmological background or of the dust radiation at the solar circle. The latter can easily increase by a factor of a few in the inner Galaxy, so synchro-self-Compton should not dominate in G21.5-0.9. The third most luminous case is found in the left wing of the Kookaburra, around PSR J1420-6048. It is fainter, but more compact. Following the same assumptions, the soft synchrotron energy density falls 25 times below the interstellar or cosmological fields, so we can consider that most of the TeV wind nebulae should mainly upscatter the cosmological background and interstellar radiation. Starlight and warm dust radiation may play a significant role below 1 TeV if the nebula resides near bright star clusters as in G0.9+0.1 [57] and G12.82-0.02 (HESS J1813-178, [50]). In any case, the GLAST sensitivity should prove very useful to constrain the peak of the inverse Compton emission, therefore the particle ageing.

Thanks to the X-ray imaging capabilities, one can often separate the diffuse nebular emission from pulsar DC

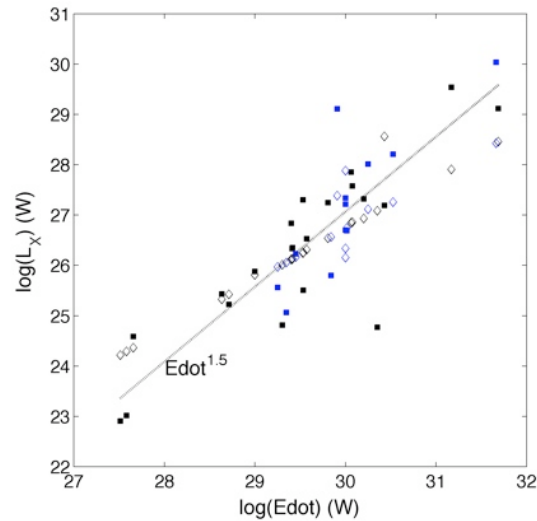

Figure 5: evolution of the 2-10 keV luminosity of pulsar wind nebulae as a function of the spindown power. The squares mark the observed values, highlighted in blue when the nebula also shines at TeV energies, and the open diamonds give the toy model predictions if 1% of the spindown power is given to the wind.

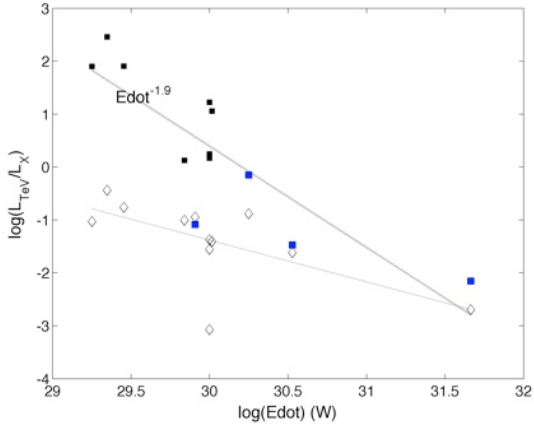

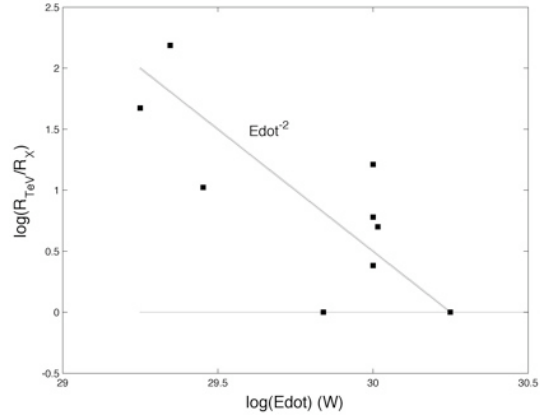

Figure 6: evolution of the ratio of the 0.5-10 TeV to the 2-10 keV luminosities of pulsar wind nebulae. The filled squares give the observed values and the open diamonds the prediction from the toy model. The three blue squares with a ratio < 0.1 are the youngest, most compact, wind nebulae in the sample (Crab, Kes 75, and G21.5-0.9). The fourth blue square marks the jet-like nebula from PSR B1509-58.

Figure 8: evolution of the size ratio between the TeV and X-ray emission regions of pulsar wind nebulae as a function of spindown power.

nebulae they had listed (details will be given in a forth-coming paper), I have plotted in Figure 5 the evolution of the 2-10 keV luminosity with $\dot{E}_{psr}$ and I have highlighted the sources that have been detected at TeV energies. The latter span a small range of ages from a thousand years in the Crab and Kes 75 to 50 kyr in PSR J1809-1917. Their X-ray luminosity is not particularly bright compared to the general scaling relation. The luminosities recorded between 0.5 and 10 TeV show a much narrower dynamical range. Except for Vela X and G21.5-0.9, which have luminosities of 8 $10^{25}$ W and 6 $10^{26}$ W, respectively, all other sources gather in the (0.3-1.4) $10^{28}$ W range and show little dependence on the pulsar spindown power, if any, and a large dispersion. This is illustrated in Figure 6 where the ratio of the TeV to X-ray luminosity scales as $L_{TeV}/L_X \propto \dot{E}_{psr}^{-1.9}$. In fact, most of the apparent evolution is due to the change in X-ray emission. This trend is nicely independent of distance.

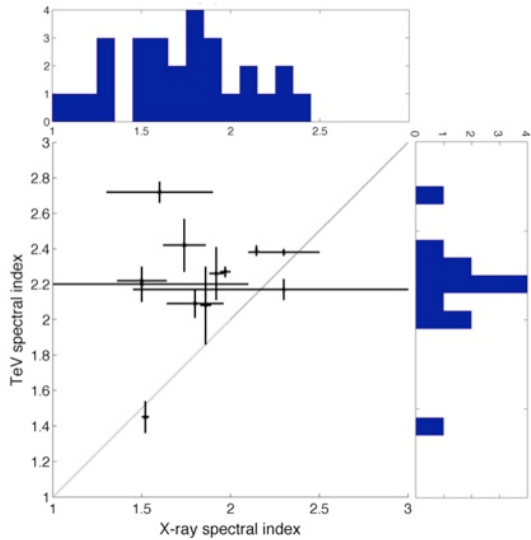

Figure 7: distributions of the photon spectral indices of the pulsar wind nebulae in X rays and at TeV energies, for the whole sample of objects displayed in Figure 5 (upper histogram) and the sub-group of TeV emitting objects (right histogram and scatter plot).

emission or the innermost part of the wind. Several authors have noted that the nebular luminosity scales with the pulsar spindown power as $L_X \propto \dot{E}_{psr}^{1.4 \pm 0.2}$ [62, and references therein]. Updating their list for the newly found nebulae and for results found in the literature on more detailed morphological studies of several

Since the TeV emitting electrons have lower energies than the X-ray ones, the evolution of the $L_{TeV}/L_X$ ratio with $\dot{E}_{psr}$ (alias youth) traces how the particles age in the nebula. Particle ageing is also illustrated by the spectral softening of the TeV emitting electrons with respect to the X-ray emitting ones. Figure 7 shows that the $\gamma$-ray spectra are often softer than the X-ray ones. Particle ageing is also reflected in the larger extent of the TeV nebula when the source is resolved in both energy bands. Figure 8 shows that the size ratio, measured in terms of radius or length depending on the source morphology, indeed increases with age (i. e. decreasing spindown power) as $R_{TeV}/R_X \propto \dot{E}_{psr}^{-2.0}$, but it presents a large dispersion. The spectral softening with distance from the pulsar strongly supports an electronic origin of the $\gamma$ radiation since synchrotron and inverse

Compton losses result in a longer lifetime at lower energy. One would expect harder spectra in γ rays than in X rays at large distance in the case of ion emission.

The higher energy electrons illuminating the X-ray synchrotron nebula give a more instantaneous image of the wind than the longer-lived electrons shining in γ rays which integrate a fair fraction of the pulsar wind history. Both the evolution of the wind power and of its magnetic field near the shock are linked to the history of the pulsar spindown power. To explore their impact on the recorded luminosities in Figures 5 and 6, I have developed a toy model based on a small set of assumptions. Given the young age (< 600 kyr) of the X-ray nebulae plotted in Figure 5, one can reasonably assume that the neutron star magnetic field has not decayed, therefore that the product $P^{(n-2)}.(dP/dt)$ of the rotational period and its time derivative remains remains constant from birth to now. n notes the pulsar braking index (n = 3 for a magnetic dipole). This implies that the spin-down power evolves as $\dot{E}_{psr}(t) = \dot{E}_0 (1 + t/\tau_0)^{(n+1)/(n-1)}$. The characteristic timescale $\tau_0$ can be retrieved from the present timing measurements and the pulsar period at birth: $\tau_0 = P_0^{(n-1)} P^{(2-n)} (n-1) / |dP/dt|$. A birth period $P_0 = 15$ ms has been adopted to account for the fastest pulsars in the sample. In the absence of pulsar timing data for the HESS J1813-178, Rabbit, and HESS J1640-465 nebulae, a spindown power of $10^{30}$ W and ages of 2, 16, and 20 kyr have been chosen respectively, following [50, 63, 64]. The pulsar powering the bow shock nebula G189.22+2.90 in IC 443 is probably older (30 kyr) and less energetic ($10^{29}$ W) [65]. A constant fraction of the spindown power $\dot{E}_{psr}(t)$ is poured into the wind at the terminal shock. It is distributed over the electron spectrum. A constant $E^{-2.1}$ spectrum has been assumed for the injected electrons above 1 GeV. I have not attempted to reproduce the trend between the X-ray photon index and the spindown power ($\alpha_X \propto \dot{E}_{psr}^{-1/2}$) that has been observed [66]. The maximum electron energy has been set to the full electrical potential drop $e\Phi_{open}$ across the open magnetosphere today. The latter evolves slowly as $\dot{E}_{psr}^{1/2}$ and has not changed much over the short lifetime of the X-ray emitting electrons, so it has been kept constant in the calculation. The particles have been traced in time, loosing energy primarily by synchrotron radiation. This should be revised in view of the large $L_{TeV}/L_X$ ratios found for middle-aged objects. Adding inverse Compton losses will remove even more particles from the γ-ray window. The magnetic energy density of the wind injected at the shock may scale as $\dot{E}_{psr}(t)$, so $B(t) \propto \dot{E}_{psr}(t)^{1/2}$. Other scaling laws have been explored ($B \propto \dot{E}_{psr}$, ...), but do not appreciably modify the results given in Figure 5 and 6. The minimum magnetic strength in the wind at birth is set to the maximum value between 30 nT and the value required so that the electrons currently injected with half the maximum energy $e\Phi_{open}$ shine at 10 keV which is the upper limit of the X-ray window. This applies only to the older pulsars that have a reduced voltage. The inverse Compton luminosity was calculated for an energy density of 0.53 MeV m$^{-3}$ of photons with an energy of 2 meV representative of the cosmological background and cold dust radiation. The most drastic assumption is that the injected particles and field remain frozen. In other words, there is no spatial evolution of the wind density or field strength in the toy model and the particles loose energy in the same initial field all their life. This is why the comparison between the predictions of the toy model and the data allows to guess what is due to the pulsar evolution and what must be due to the wind expansion and structural evolution in the data.

The results are presented for each nebula as open diamonds in Figures 5 and 6. The evolution of the predicted synchrotron luminosity in the 2-10 keV band follows reasonably well the observations given the simplicity of the toy model. The wind power has been set to 1% of the spindown power as a good match to the data. Most of the observed X-ray evolution is driven by the decrease in wind power with time because the short lifetime of the X-ray emitting electrons. They have been injected in the recent past, typically over the last 10% or 20% of the pulsar lifetime in most cases. The situation is quite different in γ rays since the emission integrates the wind history over a large fraction of the pulsar age, often more than half its age. The evolution of the predicted inverse Compton emission is much shallower than the data. The modelled $L_{TeV}/L_X$ ratio is independent of the power fraction attributed to the wind. The predicted slope in Figure 6 is rather insensitive to the initial strength as well as the evolution of the magnetic field B(t). The rapid increase of the TeV to X-ray luminosity ratio with age indicates that synchrotron losses must strongly decrease as the particles move out to keep enough of them shining in γ rays. The four nebulae in the sample for which we expect a minimal influence of the spatial evolution have been highlighted. They consist of the three very young and compact Crab, Kes 75, and G21.5-0.9 nebulae where most of the synchrotron power comes from the inner region where the magnetic pressure has built up to the equipartition value because of the wind slowing down [67]. The jet-like emission from PSR B1509-58 has also been highlighted because the comparison of the tail lengths recorded from 0.5 to 100 keV by RO-SAT, Beppo-SAX, and INTEGRAL was compatible with synchrotron ageing in a uniform magnetic field over 20 or 30 pc [56]. The toy model predictions are indeed closest to the data for these cases, especially when keeping in mind that an additional SSC component for the Crab would move the model prediction up by a factor of 3 or 4. For the others objects, the synchrotron burn-off is obviously too strong to account for the TeV flux. Adding inverse Compton losses will increase the discrepancy at old age. The model strongly under-predicts the TeV flux and it is unlikely that the ambient interstellar radiation field can be increased to match the data. So, the trend observed in Figure 6 bears valuable information on the spatial structure of the nebulae as they evolve.

The fact that the Cherenkov telescopes have detected wind nebulae at different stages of their evolution is of great interest. The three youngest objects quoted above illustrate the early development of the wind when confined by the supernova ejecta. HESS J1640-465 may bring another example when its age is known. The elongated nebular shapes in HESS J1813-178 and around PSR B1823-13, well inside their supernova remnant, and the long Vela X tail that expands almost at right angle from the pulsar spin axis give potential examples of winds crushed back by an irregular reverse shock. This may happen when the forward supernova shock slows down at different rates in azimuth because of different mass loading in a non-uniform medium [68]. Later in the evolution, when the pulsar has moved near the edge of the supernova remnant or when it has left it, its supersonic motion confines the wind inside a bow shock. The ram pressure strongly compresses the wind upfront and lets it stretch at the back. It forms a cometary tail trailing behind the pulsar. The particles injected at the front and back regions of the termination shock suffer very different losses before joining in the tail. Modelling the spatial distribution of the particle density and magnetic field strength in the crushed and bow shock configurations is difficult and being able to probe particle aging in these situations with GeV to TeV γ rays will prove very useful to constrain the models. The nebulae of PSR J1809-1917 and B1800-21 may illustrate this stage when the identification is confirmed. So will the Rabbit wind when its age is known. Understanding why the X-ray wakes of the relatively young pulsars in IC 443 and in CTB 80 (PSR B1951+32) have not been detected above 100 GeV by MAGIC also needs further investigations [69, 70], as does the case of PSR B1853+01 in W44.

Other promising candidates have been reported at GeV energies from EGRET sources and need confirmation by GLAST. The 10-kyr old, 2.2 $10^{30}$ W pulsar PSR J2229+6114 is a compelling identification for the stable 3EG J2227+6122 source that is confirmed in the revised catalogue [71]. It also coincides with a COMPTEL source in the 0.75-3 MeV band. The compact X-ray nebula, with a possible 14" jet, belongs to an incomplete non-thermal radio shell. CTA 1 should also bring an interesting case. The brightest unidentified EGRET source off the Galactic plane, 3EG J1835+5918, and its X-ray counterpart have long been proposed as a second Geminga [72, 73] because of its hard and stable spectrum cutting off at 2 GeV and because of the lack of radio and optical counterparts down to very low magnitudes. The 1.7 $10^{29}$ W compact keV nebula can be powered by a 20 kyr radio-quiet pulsar. Another promising case corresponds to the wind of PSR B1046-58. Likely γ-ray pulsations have been found in the signal from 3EGJ1048-5840, but 40 % of the emission above 400 MeV is not pulsed and a faint X-ray nebula has recently been found for this 20-kyr old, 2 $10^{29}$ W pulsar [74, 75]. The fraction of radio-quiet Geminga pulsars among the Galactic sources may be rather large if the pulsed γ-ray beams are produced at high altitude inside the magnetosphere [76]. Searching for wind nebulae with GLAST will therefore need to concentrate at energies above several GeV to benefit from the sharp cut-off expected in the pulsed emission from 10-100 kyr old neutron stars.

## Gamma-ray binaries

The detection of a pulsar wind nebula in a binary system has opened the possibility of probing the wind structure as a function of compression near periastron [77]. PSR B1259-63 indeed follows an eccentric 3.5-yr long orbit around its massive Be companion and the ram and radiation pressures build up when it crosses the equatorial outflow from the star. A variety of situations have been explored at the interface between the two winds: whether the pulsar wind is still supersonic and bounded by a termination shock at the point of pressure balance with the stellar outflow or not, or whether the Compton drag from the intense stellar radiation can slow down the unshocked wind or not. Inverse Compton components and lightcurves have been estimated [78, and references therein]. Another source of γ rays has been proposed using the termination shock of the stellar outflow against the pulsar wind to accelerate ions and electrons and letting them shine in γ rays by $\pi^0$ decay, bremsstrahlung, and inverse Compton processes [79]. The TeV detection of two other eccentric and massive binaries, LS 5039 [80] and LSI +61°303 [81], has started a debate between a pulsar wind or a microquasar jet origin of the emission. The nature of the compact object in LS 5039 and LSI +61°303 was not firmly identified and both types of systems could present striking similarities when seen from Earth. Because of the stellar brightness, a strong orbital modulation is expected in both systems from two-photon pair production [26]. It is essential to model this absorption to be able to explore the intrinsic source variability that may result from a change in accretion rate in the microquasar case or a change in wind compression in the pulsar case, and from the increase in inverse Compton emissivity near periastron in both systems. The energy distribution recorded for the three binaries were quite alike and, morphologically, comparable images could be computed from an extended jet and from a cometary wind [82]. Very high resolution radio images of LSI +61°303 showed a tail pointing away from the star along the orbit [83] which strongly suggested the presence of a bow-shock pulsar wind. So, the situation appeared confused until the detection, at the $4\sigma$ level, of the canonical black-hole binary Cygnus X-1. It was seen by MAGIC above 100 GeV for a short hour [84]. It has strengthened the case for a genuine high-energy activity from microquasars. Confirming another $\gamma$-ray flare will establish high-mass microquasars as $\gamma$-ray emitters as well as pulsar wind nebulae. Summarizing the numerous models that predict $\gamma$ radiation from microquasar jets and how they should vary with aspect angle, precession, accretion rate, and companion type is beyond the scope of this review. The studies of LS 5039 [85, 86] illustrate the various processes at work in massive systems where inverse Compton scattering of the copious stellar photons clearly dominates in the GeV to TeV band. This is probably why only massive binary systems have been detected so far. Above 1 TeV, synchro-self-Compton emission contributes a significant fraction to the overall flux. This process is the only hope of detecting low-mass binaries with current telescopes [87], unless the intensity is magnified by Doppler boosting at small viewing angles from the jet axis. We would then happily discover the long-sought microblazars.

In conclusion, with both rotation-powered and accretion powered massive binaries, isolated pulsars and the many facets of their magnetospheric and wind activities, champagne bubbles from superstar clusters and SNOBs, shock acceleration in supernova remnants, bubbles and stellar wind collisions, a wealth of GeV and TeV sources still awaiting identification, and higher performance telescopes soon to come, the future of the Galactic $\gamma$-ray club looks very bright. Let me thank the organizers for having run such a festive and lively conference where many ideas were forged and debated to make this future even brighter.